\documentclass[%
 reprint,
 amsmath,amssymb,
 aps,
]{revtex4-2}

\usepackage{mathrsfs,bm,amsmath,amssymb}
\usepackage{graphicx}
\usepackage{color}
\usepackage{hhline}
\usepackage{bm}
\def\d{{\partial}}
\def\s{{\sigma}}
\def\e{{\epsilon}}

\def\q{{ {\bm q} }}

\def\0{{ {\bm 0} }}
\def\w{{\omega}}
\def\a{{\alpha}}

\def\oddf{{\scriptsize \mbox{odd}}}
\def\evenf{{\scriptsize \mbox{even}}}
\allowdisplaybreaks[4]

\makeindex

\begin{document}

\title{
Generation of odd-frequency surface superconductivity 
with spontaneous spin current
due to the zero-energy Andreev bound state
}

\author{Shun Matsubara$^1$, Yukio Tanaka$^2$, and Hiroshi Kontani$^1$}
\affiliation{
$^1$Department of Physics, Nagoya University, Nagoya 464-8602, Japan\\
$^2$Department of Applied Physics, Nagoya University, Nagoya 464-8603, Japan
}

\date{\today}

\begin{abstract}
We propose that the odd-frequency $s$ wave
($s^\oddf$ wave) superconducting gap function,
which is usually unstable in the bulk,
naturally emerges at the edge of $d$ wave superconductors.
This prediction is based on 
the surface spin fluctuation pairing mechanism 
owing to the zero-energy surface Andreev bound state.
The interference between bulk and edge gap functions triggers the 
$d+s^\oddf$ state,
and the generated spin current is a useful signal
uncovering the ``hidden'' odd-frequency gap.
In addition, the edge $s^\oddf$ gap can be determined
via the proximity effect on the diffusive normal metal.
Furthermore, this study provides a decisive validation of
the ``Hermite odd-frequency gap function,''
which has been an open fundamental challenge to this field.
\end{abstract}

\maketitle

\section{Introduction}
In strongly correlated metals,
the introduction of an edge or interface
frequently generates new electronic states
that are quite different from the bulk ones.
For example, in unconventional or topological superconductors,
the zero-energy surface Andreev bound state (SABS) frequently emerges
and reflects the topological property of the bulk superconducting (SC) gap
\cite{Buchholtz,Nagai,Hu-ZBCP,Tanaka-ZBCP,Kashiwaya-junction,Nagato,Kashiwaya-ZBCP,Sato_ABS_topo,Kashiwaya-ZBCP-2,Alff,Wei-ZBCP,Geek-ZBCP}.
Because the flat band due to the SABS is fragile against perturbations, 
interesting symmetry-breaking phenomena 
have been actively considered theoretically
\cite{Matsumoto-Shiba-I,Matsumoto-Shiba-II,Kuboki,Fogelstrom}.
A well-known example is the edge $s$ wave state with
time-reversal-symmetry (TRS) breaking due to an attractive channel,
the so-called $d+is$ wave state
\cite{Matsumoto-Shiba-I,Matsumoto-Shiba-II}.

The huge local density of states (LDOS) in the zero-energy SABS 
also provides novel strongly correlated surface electronic states.
For example, surface ferromagnetic (FM) criticality
is naturally expected based on the Hubbard model theoretically 
\cite{Potter,Matsubara-JPSJ,Matsubara-PRB1}.
Edge-induced unconventional superconductivity would be
one of the most interesting phenomena due to FM criticality.
Based on this mechanism,
two of the present authors previously proposed an edge-induced $p$ wave SC state 
on $d$ wave superconductors
\cite{Matsubara-PRB2}.
Another exotic possibility of the edge SC state is the  
``odd-frequency SC state.''
However, regardless of the difficulties in its realization,
the odd-frequency SC state
is attracting considerable attention in the field of superconductivity
because the varieties of pairing symmetry are doubled by 
allowing the odd parity with respect to time
\cite{Berezinskii,Coleman,Kirkpatrick,Balatsky-Abrahama,Tanaka-odd-frequency,Bala-rev}.
Therefore, an accessible method for generating the odd-frequency gap function
is proposed in this study.
Although it is possible to consider 
the induced odd-frequency gap function near the edge \cite{Matsumoto2013}, 
there has not been microscopic theory  in realistic systems. 

The mechanisms and properties of the odd-frequency SC states
have been actively discussed by many theorists
\cite{Berezinskii,Coleman,Kirkpatrick,Balatsky-Abrahama,Matsumoto2013,Fominov,Yada,Fuseya,Solenov,Kusunose,Kusunose2}.
Based on the spin-fluctuation theory, FM (antiferromagnetic) 
fluctuations can mediate
odd-frequency superconductivity with the $s$ wave triplet ($p$ wave singlet) gap 
\cite{Bulut,Vojta,Fuseya,Hotta,Shigeta,Yanagi}.
However, 
if the odd-frequency gap function is Hermitian,
it is unstable as a bulk state due to the inevitable emergence of 
the ``paramagnetic Meissner'' (para-Meissner) effect 
\cite{Balatsky-Abrahama,Kirkpatrick,Heid}.
To escape from this difficulty, inhomogeneous SC states
with a large center of mass momentum
of the gap function have been considered
\cite{Hoshino}.
In contrast, a homogeneous
non-Hermitian odd-frequency gap function with 
the usual Meissner effect has been proposed
\cite{Solenov,Kusunose,Kusunose2}.
However, mixing between Hermitian and non-Hermitian
odd-frequency ``pair amplitudes'' gives rise to an unphysical
imaginary contribution to the Josephson current and superfluid density
\cite{Fominov}.
At present, the essential properties of the 
odd-frequency gap function remain unknown.
To address this challenge, it would be beneficial to study the
coexisting states of the odd-frequency 
and well-known even-frequency gap functions.

In this study, we predict that the odd-frequency spin-triplet $s$ wave 
($s^\oddf$ wave)
gap function naturally emerges at the edge of $d$ wave superconductors,
which is mediated by SABS-induced magnetic fluctuations
\cite{Matsubara-JPSJ,Matsubara-PRB1,Matsubara-PRB2}.
This prediction is derived from the analysis of the edge SC gap equation
based on the cluster Hubbard model with the bulk $d$ wave gap.
The obtained bulk+edge superconductivity with TRS
accompanies the spontaneous edge spin current, which is
an important signal for determining the ``hidden'' odd-frequency SC gap.
This study provides a decisive validation of
the {\it spatially localized} odd-frequency gap function 
with the para-Meissner effect.

It is known that the odd-frequency pair amplitude can be induced by external symmetry breaking
from conventional even-frequency pairing.
If spin-rotational symmetry is broken,
odd-frequency spin-triplet $s$ wave pairing can be induced from the conventional spin-singlet one shown in
the superconductor/ferromagnet junction
\cite{Bergeret_2001,Bergeret_2005,Golubov_2002,Yokoyama_2007,
Buzdin_2005,Buzdin_2011,Buzdin_2012,Linder_2009,Halterman_2014,Eschrig_2003,Asano_2007,Cayao_2020}.
On the other hand, translational symmetry breaking can also induce odd-frequency pairing
from bulk even-frequency superconductors with translational symmetry braking
\cite{Tanaka-odd-frequency,Tanaka-odd-frequency-2,Tamura,Tanaka_2007_1,Tanaka_2007_2}.
In this case, spin-singlet odd-parity (spin-triplet even-parity) pairing can be generated
from spin-singlet even-parity (spin-triplet odd-parity) bulk superconductors
\cite{Tanaka-odd-frequency,Tanaka-odd-frequency-2,Tamura,Tanaka_2007_1,Tanaka_2007_2}.
In these cases,
the anomalous proximity
\cite{Tanaka2004,Higashitani2009}
and para-Meissner effects 
\cite{Higashitani,Tanaka2005R,Suzuki-Asano1,Suzuki-Asano2,Bernardo,Krieger}
are induced even if the $s^\oddf$ wave gap function is zero.
In particular, the odd-frequency amplitude is enlarged by the zero-energy SABS,
and it can induce the $s^\oddf$ wave gap function via the $U$ introduced
in this study.
In this paper, we study the emergence of the odd-frequency superconducting gap function 
in the paramagnetic state, mediated by edge-induced ferromagnetic fluctuations. 

To investigate the strong correlation effects induced by the 
huge LDOS in the edge SABS,
we apply spin-fluctuation theory
\cite{Moriya,Scalapino,Tremblay,Chubukov,Kontani-rev}
to the cluster Hubbard model with an edge structure, as illustrated
in Fig. \ref{fig:fig1} (a). 
This framework is useful for electronic systems without periodicity
because it can naturally elucidate the impurity-induced 
enhancement of the magnetic fluctuations 
observed in cuprate superconductors
\cite{Chen,Kontani-imp,Alloul99-2,Ishida96}.
Note that the non-Fermi liquid transport phenomena and 
$d$ wave bond order in cuprates are well understood 
based on the spin-fluctuation theories
\cite{Moriya,Kontani-rev},
by considering vertex corrections correctly
\cite{Kontani-rev,Yamakawa-CDW,Kawaguchi-CDW,Tsuchiizu4}.

\section{Model and Theoretical Method}

\begin{figure}[htb]
\includegraphics[width=.92\linewidth]{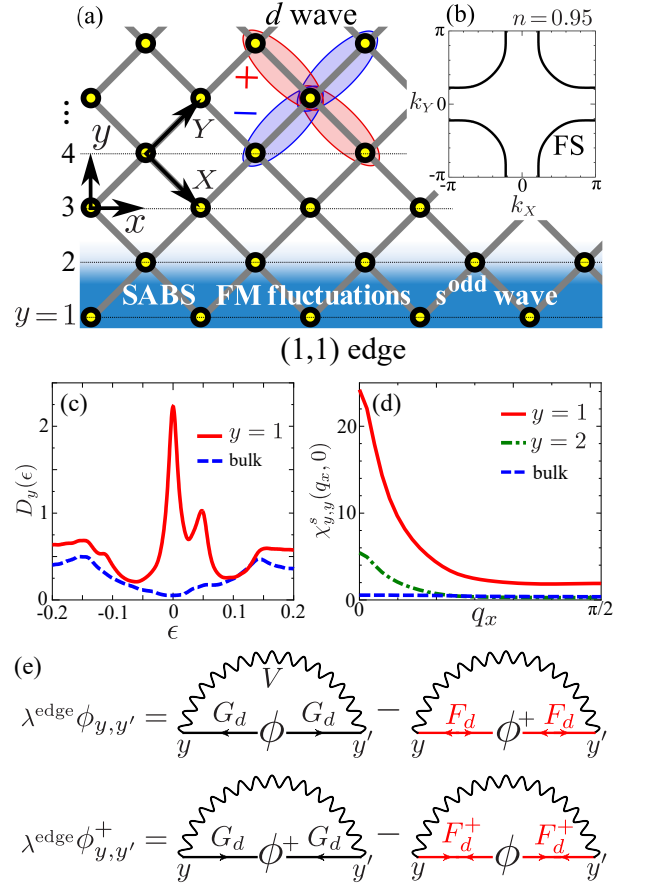}
\caption{
(a) Cluster Hubbard model with a (1,1) edge.
The orthogonal unit vectors ($\hat{\bm{x}}$, $\hat{\bm{y}}$)
and ($\hat{\bm{X}},\hat{\bm{Y}}$) are illustrated.
(b) Bulk FS.
(c) SABS-induced peak in the LDOS at $\Delta^d=0.16$
in the case with quasiparticle damping $\gamma=0.01$.
(d) Edge-induced FM fluctuations obtained via the 
site-dependent RPA $\chi^s_{y,y}(q_x,0)$ 
for $T=0.05$ and $\Delta^d=0.10$.
(e) Linearized edge gap ($\phi^{(+)}$) equation composed of 
Green's functions, $G_d$ and $F_d$, for $\Delta^d\ne0$.
$\lambda^{\rm edge}$ is the eigenvalue.
The second terms on the right-hand side determine the 
phase difference between $\Delta^d$ and $\phi$ ($\phi^{+})$.
}
\label{fig:fig1}
\end{figure}

The Hamiltonian is expressed as:
\begin{eqnarray}
\mathcal{H}=H_0
+U\sum_{i}n_{i\uparrow}n_{i\downarrow}
+\sum_{i,j}
\Delta_{i,j}^d\left(c_{i\uparrow}^\dagger c_{j\downarrow}^\dagger 
+ {\rm H.c} \right),
\label{eqn:Hamiltonian}
\end{eqnarray}
where 
$U$ denotes the on-site Coulomb interaction.
$H_0=\sum_{i,j,\sigma}t_{i,j}c_{i\sigma}^\dagger c_{j\sigma}$
represents the kinetic term, where 
$t_{i,j}$ denotes the hopping integral between sites $i$ and $j$.
In this study, we set $(t_1,t_2,t_3)=(-1,1/6,-1/5)$, 
where $t_n$ is the $n$-th nearest neighbor hopping integral
and it corresponds to the ${\rm Y}{\rm Ba}_2{\rm Cu}_3{\rm O}_{7-x}$ (YBCO) model
\cite{Matsubara-JPSJ,Matsubara-PRB1,Matsubara-PRB2,Kontani-rev}. 
The energy unit is $|t_1|=1$.
The Fermi surface (FS) in the periodic system
is illustrated in Fig. \ref{fig:fig1} (b).
$\Delta_{i,j}^d$ is the bulk $d_{xy}$ wave ($d_{X^2-Y^2}$ wave)
gap function given as
$\Delta_{i,j}^d=(\Delta^d/4)(\delta_{\bm{r}_i-\bm{r}_j,\pm \hat{\bm{X}}}
-\delta_{\bm{r}_i-\bm{r}_j,\pm \hat{\bm{Y}}})$.
A similar bulk $d$ wave gap function is microscopically 
obtained based on spin-fluctuation theories.
Considering this fact,
we introduce $\Delta^d$ as the model parameter
to simplify the discussion.
To reproduce the suppression of the $d$ wave gap near the edge,
we multiplied the $d$ wave gap function by the decay factor 
$\{1-\exp[(y_i+y_j-2)/2\xi_d]\}$
\cite{Matsubara-PRB2}.
Then, we set the coherence length $\xi_d=10$.
Figure \ref{fig:fig1} (c) presents the LDOS at the edge site
for $\Delta^d=0.16$.
The obtained sharp SABS-induced peak in LDOS drives the system 
towards a strong correlation
\cite{Matsubara-PRB1}.
In the following numerical study, we set the filling as $n=0.95$.
The numerical results are essentially unchanged for $n=0.8$--$1.2$.

In this study, we introduce the $2N_y\times 2N_y$ Nambu Green's function
in the presence of the bulk $d$ wave gap
$\Delta_{y,y'}^{d}(k_x)\equiv{\Delta^{d}}^{\uparrow\downarrow}_{y,y'}(k_x)$.
Since we assume that $\Delta_{i,j}^d$ is real,
$\left\{\Delta_{y',y}^{d}(k_x)\right\}^*=\Delta_{y,y'}^{d}(k_x)$ is satisfied.
Thus, we consider the following Nambu Hamiltonian
\cite{Matsubara-PRB1,Matsubara-PRB2}:
\begin{align}
\mathcal{H}_{d}
&=
\sum_{k_x}
\left(
^{t}\hat{c}_{k_x,\uparrow}^{\dag},
^{t}\hat{c}_{-k_x,\downarrow}
\right)
\left(
    \begin{array}{cc}
\hat{H}^{0}(k_x)           &     \hat{\Delta}^{d}(k_x) \\
\hat{\Delta}^{d}(k_x)    &      -^{t}\hat{H}^{0}(-k_x)\\
    \end{array}
\right)
\nonumber\\
&~~~~\times
\left(
    \begin{array}{c}
      \hat{c}_{k_x,\uparrow} \\
      \hat{c}_{-k_x,\downarrow}^{\dag} \\
    \end{array}
  \right),
\label{eq:nanbu_2}
\end{align}
where $\hat{c}_{k_x,\uparrow}$ and $\hat{c}_{-k_x,\downarrow}^{\dag}$ represent
the $N_y$-component column vector of sites.
Next, we define the Green's functions in the bulk $d$ wave SC state as follows:
\begin{eqnarray}
& &
  \left(
    \begin{array}{cc}
     \hat{G}_{d}(k_x,i\epsilon_n) & \hat{F}_{d}(k_x,i\epsilon_n) \\
     \hat{F}^{+}_{d}(k_x,i\epsilon_n) & -^{t}\hat{G}_{d}(-k_x,-i\epsilon_n)  
    \end{array}
  \right)
\nonumber\\
& &
=
  \left(
    \begin{array}{cc}
     i\epsilon_n\hat{1}-\hat{H}^0(k_x) & -{\hat{\Delta}}^{d}(k_x) \\
     -{\hat{\Delta}}^{d}(k_x) & i\epsilon_n\hat{1}+^{t}\hat{H}^0(-k_x)
    \end{array}
  \right)^{-1}.
\label{eq:sc-4}
\end{eqnarray}
Then, we calculate the site-dependent spin susceptibility
$\chi^s_{y,y'}(q_x,i\w_l)$ in the cluster Hubbard model 
with the bulk $d$ wave gap in Eq. (\ref{eqn:Hamiltonian}),
using the real-space random-phase-approximation (RPA).
Here, we adopt the $k_x$ representation
by considering the translational symmetry,
and $\w_l=2\pi T l$ represents the boson Matsubara frequency.
The irreducible susceptibilities are given by $\hat{G}_{d}$, $\hat{F}_{d}$, and $\hat{F}_{d}^{+}$ as
\begin{eqnarray}
\chi^0_{y,y'}({q}_x,i\omega_l) &=&-T\sum_{{k}_x,n}
{G_d}_{y,y'}({q}_x+{k}_x,i\omega_l+i\epsilon_n)
\nonumber\\
&&\times {G_d}_{y',y}({k}_x,i\epsilon_n) ,
\label{eqn:chi0}
\end{eqnarray}
\begin{eqnarray}
\varphi^0_{y,y'}(q_x,i\omega_l)
&=&
-T
\sum_{k_x,n}
{F_d}_{y,y'}(q_x+k_x,i\omega_l+i\epsilon_n)
\nonumber\\
&&\times
{F_d}_{y',y}^{+}(k_x,i\epsilon_n).
\label{eqn:phi0}
\end{eqnarray}
$\varphi^0$ is finite only in the SC state.
The $N_y\times N_y$ matrix of the spin (charge) susceptibility $\hat{\chi}^{s(c)}$
is calculated using $\hat{\chi}^0$ and $\hat{\varphi}^0$ as
\begin{eqnarray}
&&
\hat{\chi}^{0s(c)}(q_x,i\omega_l)
=
\hat{\chi}^0(q_x,i\omega_l)+(-)\hat{\varphi}^0(q_x,i\omega_l),
\label{eqn:phisc}
\end{eqnarray}
\begin{eqnarray}
\hat{\chi}^{s(c)}(q_x,i\omega_l)
&=&
\hat{\chi}^{0s(c)}(q_x,i\omega_l)
\nonumber\\
&&
\times
\left\{
\hat{1}-(+)U
\hat{\chi}^{0s(c)}(q_x,i\omega_l)
\right\}^{-1}.
\label{eqn:chis}
\end{eqnarray}
The spin Stoner factor is the largest eigenvalue of 
$U\hat{\chi}^{0s}(q_x,i\omega_l)$ at $\omega_l=0$.
The magnetic order is realized when $\alpha_S\geq1$.
The pairing interaction for the triplet SC is given by
\begin{eqnarray}
\hat{V}({q}_x,i\omega_l)
=U^2\left(-\frac12\hat{\chi}^s({q}_x,i\omega_l)
-\frac12\hat{\chi}^c({q}_x,i\omega_l)\right).
\nonumber\\
\label{eq:trip_gap_int}
\end{eqnarray}
Figure \ref{fig:fig1}(d) illustrates the obtained $\chi^s_{y,y}(q_x,0)$
in the $y$th layer at zero frequency.
The obtained strong FM fluctuations ($q_x\approx0$)
originate from the SABS 
\cite{Matsubara-PRB1},
and they mediate the spin-triplet edge-induced superconductivity
\cite{Matsubara-PRB2}.

The linearized triplet gap equations for
$\hat{\phi}(k_x,i\e_n) \ (\propto \langle c_{k_x\uparrow}c_{-k_x\downarrow}\rangle)$ 
and $\hat{\phi}^+(k_x,i\e_n) \ (\propto \langle c_{-k_x\downarrow}^\dagger c_{k_x\uparrow}^\dagger\rangle)$ 
are presented in Fig. \ref{fig:fig1}(e),
and their analytic expressions are
\begin{eqnarray}
& &\lambda^{\rm edge}
\phi_{y,y'}(k_x,i\epsilon_n)
\nonumber\\
&=&
-
T
\sum_{k_x',Y,Y',m}
V_{y,y'}(k_x-k_x',i\epsilon_n-i\epsilon_m)
\nonumber\\
& &
\times
\left\{
G_{y,Y}(k_x',i\epsilon_m)
\phi_{Y,Y'}(k_x',i\epsilon_m)
G_{y',Y'}(-k_x',-i\epsilon_m)
\right.
\nonumber\\
& &
-
\left.
F_{y,Y}(k_x',i\epsilon_m)
{\phi^{+}_{Y,Y'}}(k_x',i\epsilon_m)
F_{Y',y'}(k_x',i\epsilon_m)
\right\},
\label{eq:trip_gap_eq1}
\end{eqnarray}
\begin{eqnarray}
& &\lambda^{\rm edge}
{\phi^{+}_{y,y'}}(k_x,i\epsilon_n)
\nonumber\\
&=&
-
T
\sum_{k_x',Y,Y',m}
V_{y,y'}(k_x-k_x',i\epsilon_n-i\epsilon_m)
\nonumber\\
& &
\times
\left\{
G_{Y,y}(-k_x',-i\epsilon_m)
{\phi^{+}_{Y,Y'}}(k_x,i\epsilon_m')
G_{Y',y'}(k_x',i\epsilon_m)
\right.
\nonumber\\
& &
-
\left.
F_{y,Y}^{+}(k_x',i\epsilon_m)
\phi_{Y,Y'}(k_x',i\epsilon_m)
F_{Y',y'}^{+}(k_x',i\epsilon_m)
\right\}.
\label{eq:trip_gap_eq2}
\end{eqnarray}
(We did not study the singlet gap equation 
because FM fluctuations suppress spin-singlet gaps.)
Because the spin-orbit interaction was absent,
we assumed that $\bm{d}\parallel \bm{z}$ ($S_z^{\rm triplet}=0$)
in the triplet gap without the loss of generality.
A detailed derivation is presented in Appendix A.
Based on Ref. \cite{Matsubara-PRB2}, 
we derived the even-frequency $p$ wave triplet gap 
$\hat{\phi}(k_x,i\e_n)=\hat{\phi}(k_x,-i\e_n)$,
where $\e_n=(2n+1)\pi T$.
However, this is not a unique possibility
because the odd-frequency pairing state 
$\hat{\phi}(k_x,i\e_n)=-\hat{\phi}(k_x,-i\e_n)$ is not prohibited in principle.

\section{Numerical Results}

\begin{figure}[htb]
\includegraphics[width=.99\linewidth]{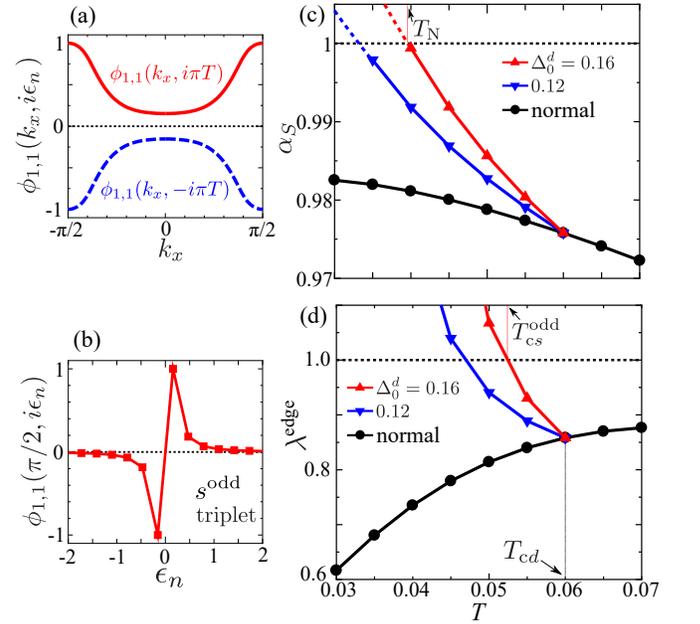}
\caption{
Obtained $s^\oddf$ wave triplet gap at the edge:
(a) $\phi_{1,1}(k_x,\pm i\pi T)$ in the first BZ $(-\pi/2<k_x\le\pi/2)$
and (b) $\phi_{1,1}(k_x=\pi/2,i\e_n)$ in the $\Delta_0^d=0.16$ case at $T=0.05$.
Obtained $T$-dependences of (c) the Stoner factor $\a_S$ and
(d) the eigenvalue $\lambda^{\rm edge}$ 
for the $s^\oddf$ wave state.
Here, the bulk $d$ wave SC gap appears at $T_{{\rm c}d}=0.06$.
In addition, $2\Delta_0^d/T_{{\rm c}d}= 4.0$--$5.3$ for 
$\Delta_0^d=0.12$--$0.16$.
The edge $s^\oddf$ wave gap is obtained for 
$\a_S \gtrsim 0.95$ at $T=T_{{\rm c}d}$.
}
\label{fig:fig2}
\end{figure}

In the triplet state,
the even-frequency (odd-frequency) gap exhibits an odd (even) parity in space
due to fermion anticommutation relations.
Considering both possibilities equally,
we analyze the gap equation in Fig. \ref{fig:fig1} (e) 
by considering the $i\e_n$ dependence of $\hat{\phi}(k_x,i\e_n)$ comprehensively.
Here, we assume the Hermitian odd-frequency gap function
\cite{Fominov,Matsumoto2013}:
\begin{eqnarray}
\phi_{y,y'}^+(k_x,i\e_n)=[\phi_{y',y}(k_x,-i\e_n)]^*
\label{eqn:Fom}
\end{eqnarray}
The reliability of this relationship will be clarified later.
We assumed the BCS-type bulk gap function 
$\Delta^d(T)=\Delta^d_0 \tanh(1.74\sqrt{T_{{\rm c}d}/T -1})$
with the transition temperature $T_{{\rm c}d}=0.06$,
which corresponds to $\sim100$ K in cuprates
for $z|t_1|\sim 1500$ K with $z=m/m^*\sim0.3$.
Experimentally, $4<2\Delta_0^d/T_{{\rm c}d}<10$ in YBCO \cite{cuprate_coherence_1,cuprate_coherence_2}.
Thus, we set $\Delta_0^d=0.12$ or $0.16$, which corresponds to $2\Delta_0^d/T_{{\rm c}d}= 4.0$--$5.3$.
We set $U=2.32$, where
the spin Stoner factor $\a_S$ is $0.975$ at $T=T_{{\rm c}d}$.

\subsection{$s^{\rm odd}$ wave SC state}
Figures \ref{fig:fig2}(a) and 2(b) exhibit the 
$k_x$ and $i\e_n$ dependences of the odd-frequency $s$ wave
($s^\oddf$ wave) gap 
for $\Delta_0^d=0.16$ at $T=0.05$, respectively.
Here, the odd-frequency $s^\oddf$ wave state 
is obtained as the largest eigenvalue state.
At the edge, the pure $s^\oddf$ gap function is obtained 
because the $d$ wave gap is zero at $y=1$.

Figures \ref{fig:fig2}(c) and 2(d) exhibit the obtained 
spin Stoner factor $\a_S$ and the eigenvalue $\lambda^{\rm edge}$
as a function of $T$, respectively.
Since SC susceptibility is proportional to 
$1/|1-\lambda^{\rm edge}|$,
the edge-gap function is expected to appear
when $\lambda^{\rm edge}\sim1$.
In the normal state ($\Delta_0^d=0$), 
$\lambda^{\rm edge}$ decreases at low $T$ 
because the pairing interaction for the odd-frequency SC gap
is proportional to $T\chi^s(\q_x,0)\propto T/(1-\a_S)$
\cite{Bulut,Vojta,Fuseya,Hotta,Shigeta,Yanagi}.
This is a well-known difficulty of the spin-fluctuation-mediated
odd-frequency SC mechanism in bulk systems.
In contrast, in the presence of the SABS,
$\a_S$ increases rapidly due to the 
huge LDOS at zero energy
\cite{Chen,Matsubara-PRB1}.
Therefore, $\lambda^{\rm edge}$ rapidly approaches unity
owing to the SABS-induced magnetic criticality
\cite{Matsubara-PRB2}.
Thus, the SABS-driven odd-frequency SC mechanism 
is naturally realized at the edge of $d$ wave superconductors.

\begin{figure}[htb]
\includegraphics[width=.99\linewidth]{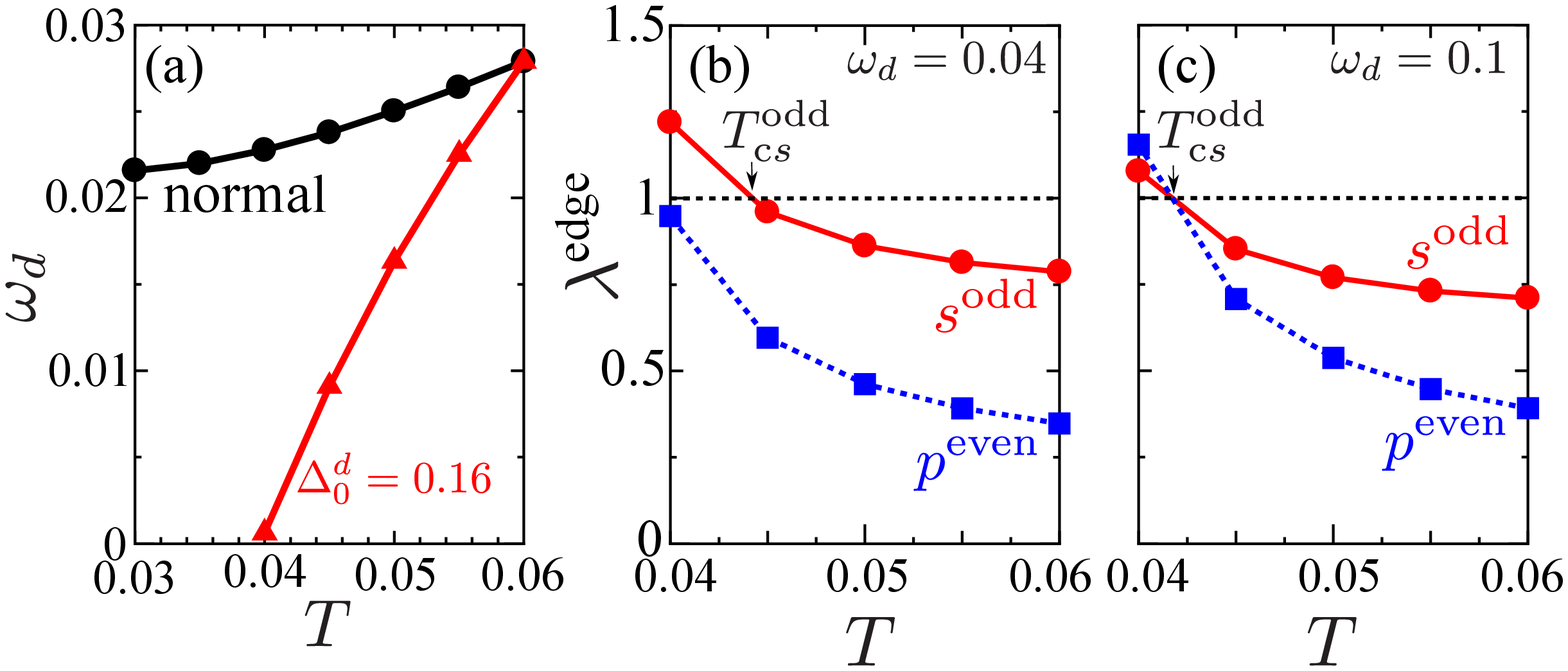}
\caption{
(a) Obtained energy-scale of the dynamical spin susceptibility 
$\w_d \ (\propto 1-\a_S)$ as a function of $T$.
Eigenvalues $\lambda^{\rm edge}$ obtained
by the pairing interaction ${\hat \chi}^s(q_x,0)\Omega(\w_l;\w_d)$
for (b) $\w_d=0.04$ and (c) $0.1$.
As it approaches the magnetic criticality $\w_d\rightarrow0$, 
$T_{{\rm c}s}^\oddf$ increases whereas
$T_{{\rm c}p}^\evenf$ decreases.
In (b), $T_{{\rm c}s}^\oddf$ is higher than $T_{{\rm c}p}^\evenf$.
}
\label{fig:fig3}
\end{figure}

\subsection{$s^{\oddf}$ wave SC state dominates $p^\evenf$ wave SC state}
Here, we discuss the reason behind the
edge $s^\oddf$ wave state dominating the edge
even-frequency $p^\evenf$ wave state in this study.
In the $k_x$-space argument,
the larger condensation energy is expected 
in the nodeless $s^\oddf$ wave state.
In the $\e_n$-space argument, proximity
to the magnetic criticality ($\a_S\lesssim1$) is crucial:
The edge pairing interaction 
$V_{1,1}(q_x,i\w_l)\propto \chi^s_{1,1}(q_x,i\w_l)$
at $q_x\sim 0$ is well fitted by the function
$\Omega(\w_l;\w_d)= \w_d/(|\w_l|+\w_d)$, 
and the obtained $\w_d$ in the present real-space RPA study
is presented in Fig. \ref{fig:fig3} (a).
$\w_d \ (\propto 1-\a_S)$ 
approaches zero at the magnetic critical point, 
and the eigenvalues of even- and odd-frequency solutions
become similar
\cite{Bulut,Vojta,Fuseya,Hotta,Shigeta,Yanagi}.
To verify this discussion, 
we compare the eigenvalues $\lambda^{\rm edge}$ of both $s^\oddf$ wave 
and $p^\evenf$ wave states,
by introducing a separable pairing interaction
$V_{y,y'}(q_x,i\w_l)\propto \chi^s_{y,y'}(q_x,0)\Omega(\w_l;\w_d)$.
The obtained results are presented in Figs. \ref{fig:fig3}(b) and \ref{fig:fig3}(c)
for $\w_d=0.04$ and $\w_d=0.1$, respectively.
It is verified that
the $s^\oddf$ wave dominates the $p^\evenf$ wave
near the quantum criticality $\w_d=0.04$,
which corresponds to the RPA
study demonstrated in Fig. \ref{fig:fig2}.
The obtained $s^\oddf$ wave state should be 
robust against impurity scattering according to the Anderson theorem.

The obtained edge $s^\oddf$ wave gap in the $\e_n$ representation is real
in the case of $\Delta^d={\rm real}$.
That is, $\phi_{1,1}(k_x,i\e_n) \propto \e_n$ is real for small $\e_n$.
Then, after the analytic continuation, 
$\phi'=[\phi_{1,1}^{\rm R}(k_x,\e)+\phi_{1,1}^{\rm A}(k_x,\e)]/2\propto i\e$ 
becomes purely imaginary.
In addition, the triplet gap function is odd with respect to the time reversal.
Therefore, the obtained state is the TRS ``$d+s^\oddf$ wave state.''
Because
$\phi''=[\phi_{1,1}^{\rm R}(k_x,0)-\phi_{1,1}^{\rm A}(k_x,0)]/2$
also approaches zero near the magnetic criticality
\cite{Fuseya},
the edge $s^\oddf$ wave gap will not affect the LDOS at zero-energy.
This result is consistent with the ubiquitous presence of 
the zero-bias conductance peak in the tunneling spectroscopy of  
cuprates \cite{Alff,Wei-ZBCP,H.Kashiwaya,Bouscher}

\begin{figure}[htb]
\includegraphics[width=.85\linewidth]{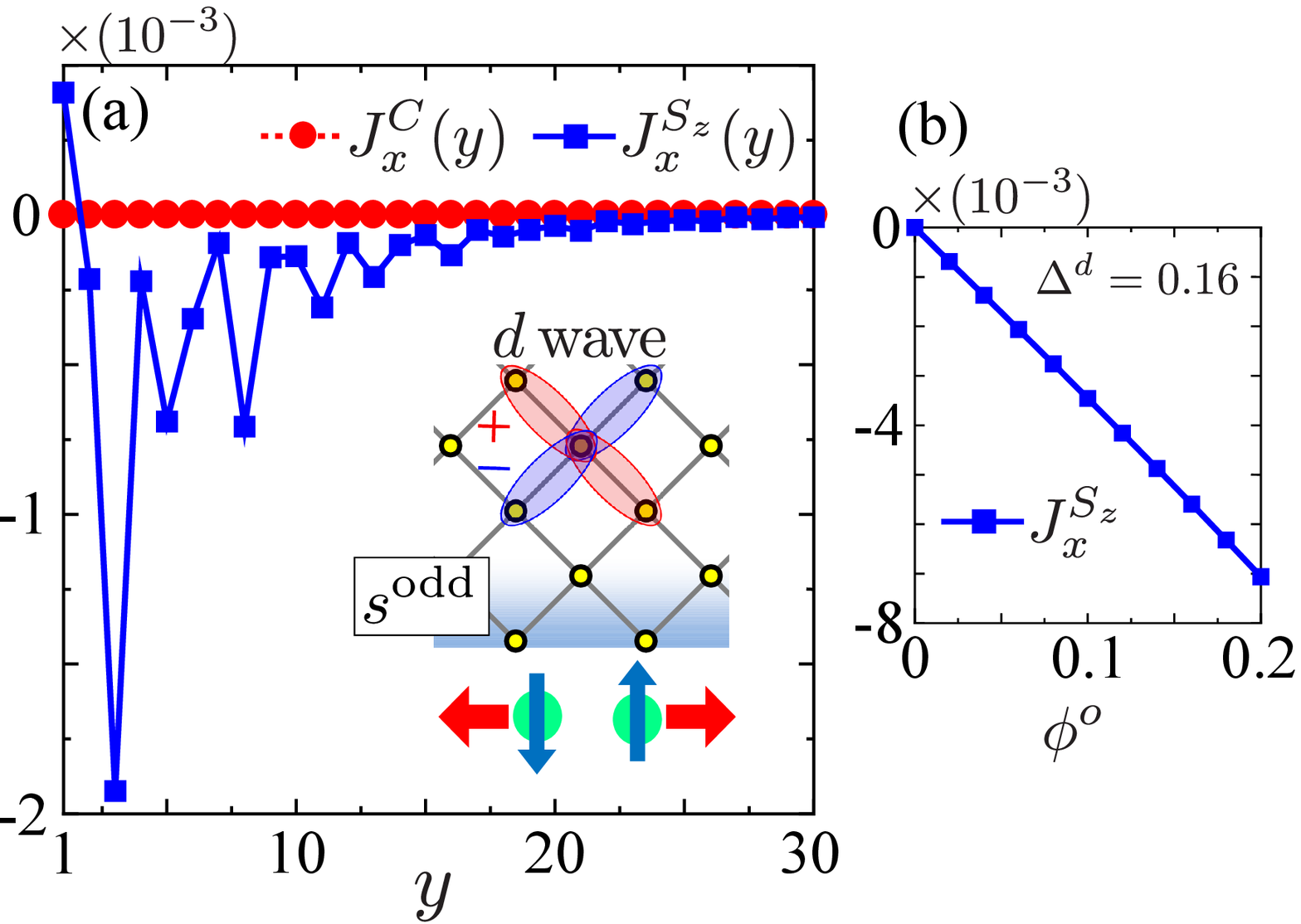}
\caption{
(a) Obtained edge currents in the $d+s^\oddf$ state
derived from the edge gap equation shown in Fig. \ref{fig:fig1} (e).
Here, $d=d_{xy}$.
The edge currents in the $p+is^\oddf$, $d+is^\evenf$, and
$p+s^\evenf$ states are illustrated in Appendix D
and listed in Table \ref{tab:table1}.
Here, we set $\Delta^{d}=0.16$ while
the $s^\oddf$ wave gap function is set as
$\phi_{y,y'}(i\e_n)= \phi^{\rm o} f^{\rm o}(\e_n)\delta_{y,1}\delta_{y',1}$
with $\phi^{\rm o}=0.16$,
where $f^{\rm o}(\e_n)$ is given in Fig. \ref{fig:fig2} (b).
(b) Obtained total edge current $J_x^{{\rm S}z}$ for $\Delta^d=0.16$
as a function of $\phi^{\rm o}$.
}
\label{fig:fig4}
\end{figure}

\subsection{Edge super current}
Here, we elucidate the emergence of the nontrivial 
edge supercurrent in the $d+s^\oddf$ wave state.
In the present cluster model with the $d+s^\oddf$ wave gap, 
the charge current along the $x$ axis from layer $y$ 
[Fig. \ref{fig:fig1}(a)] to any layer is calculated as
\begin{eqnarray}
J_x^{\rm C}(y)&=&\sum_{k_x,y',\sigma,\rho}
\big\{
(-e \delta_{\s,\rho}) v_x(k_x,y,y')
\nonumber \\
& &\times \mathcal{G}_{y',y}^{\s,\rho}(k_x,i\e_n)e^{-i\e_n0}
+(y\leftrightarrow y')
\big\},
\label{eq:charge_current}
\end{eqnarray}
where 
$v_x(k_x,y,y')\equiv \d H_{y,y'}^0(k_x)/\d k_x$
\cite{PALee}
and $\mathcal{G}_{y',y}^{\s,\rho}$ presents the Green's function for the $d+s^\oddf$ state
in Appendix A.
Here, we set
$\phi_{y,y'}(i\e_n)= \phi^{\rm o} f^{\rm o}(\e_n)\delta_{y,1}\delta_{y',1}$,
with $\phi^{\rm o}=\Delta_0^d=0.16$,
where $f^{\rm o}(\e_n)$ is provided in Fig. \ref{fig:fig2} (b).
The numerical results obtained are insensitive to the parameters 
$\phi^{\rm o}$ and $\Delta_0^d=0.16$.
Accordingly, the total edge current is $J_x^{\rm C}=\sum_yJ_x^{\rm C}(y)$.
We also calculate the spin current along the $x$ axis $J_x^{{\rm S}\mu}(y)$,
where $\mu$ represents the spin current polarization.
It is obtained by replacing 
$(-e\delta_{\s,\rho})$ with $(\hbar\hat{\s}^\mu_{\s,\rho})$
in Eq. (\ref{eq:charge_current}), where 
$\hat{\s}^\mu$ depicts the Pauli matrix.
Because $s_z$ is conserved in the present SC state, 
$J_x^{{\rm S}\mu}(y)$ is zero for $\mu=x,y$.
We emphasize that $J_x^{{\rm S}\mu}(y)$ remains constant under the time reversal.

Figure \ref{fig:fig4}(a) presents the obtained currents 
in the $d+s^\oddf$ wave state by setting $e=\hbar=1$.
Here, the charge current $J_x^{\rm C}(y)$ vanishes identically,
which is consistent with the experimental reports of muon spin rotation ($\mu$-SR)
\cite{Saadaoui}; however, the
non zero spin current $J_x^{{\rm S}z}(y)$ flows spontaneously.
The spin current polarization is parallel to the $\bm{d}$ vector.
Here, the parity of the mirror operation ${\cal M}_x$ is odd 
because the $d_{xy}$ ($s^\oddf$) gap has odd (even) parity.
In addition, the spin exchange parity is $-1$. 
Consequently, conduction electrons acquire
spin-dependent velocity, and therefore, $J_x^{{\rm S}z}(y)\ne0$.
The obtained total spin current
$J_x^{{\rm S}z}\equiv \sum_{y}J_x^{{\rm S}z}(y)$
is $\phi^{\rm o}$ linear, as shown in Fig. \ref{fig:fig4} (b).
Because $J_x^{{\rm S}z}$ is linear in $|\phi^{\rm o}|$,
a sizable amount of spin current is expected.

\begin{figure}[htb]
\includegraphics[width=.99\linewidth]{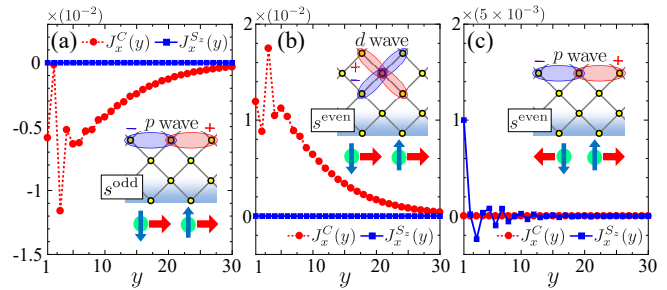}
\caption{
Obtained edge currents in
(a) the $p+is^\oddf$ wave state,
(b) the $d+is^\evenf$ wave state, and 
(c) the $p+s^\evenf$ wave state.
Here, $p=p_x$, $d=d_{xy}$, and $\Delta^{p,d}=0.16$.
We set the $s^\oddf$ wave gap function as
$\phi_{y,y'}(i\e_n)= \phi^{\rm o} f^{\rm o}(\e_n)\delta_{y,1}\delta_{y',1}$
with $\phi^{\rm o}=0.16$,
where $f^{\rm o}(\e_n)$ is given in Fig. 2(b).
We also set the $s^\evenf$ wave gap 
$\phi_{y,y'}(i\e_n)= \phi^{\rm e} \delta_{y,1}\delta_{y',1}$ 
with $\phi^{\rm e}=0.16$.
}
\label{fig:fig5}
\end{figure}

Furthermore,
we also study the edge currents
in the $d+is^\evenf$, $p+is^\oddf$, and $p+s^\evenf$ wave states.
In the TRS breaking $p+is^\oddf$ wave state ($p=p_x$),
we find that the finite charge current emerges
as shown in Fig. \ref{fig:fig5} (a),
whereas spin current vanishes.
In the $p+is^{\oddf}$ wave state,
the parity of ${\cal M}_x$ is odd,
while the parity of the spin part is even.
As a result, $J_x^{\rm C}\ne0$ is realized.
The present study is a nontrivial extension of the theory of the
$d+is^\evenf$ wave state \cite{Matsumoto-Shiba-II}.

We notice that, when the bulk SC gap is $p_x$ wave,
the SABS that drives the edge 
$s^\oddf$ wave state is absent
\cite{Kashiwaya-ZBCP}.
In the $p_X$ wave SC state,
the SABS exists, and the parity of ${\cal M}_x$ is not completely even.
Therefore, the $p_X$ wave SC state is favorable to realize
the odd-frequency SC state with a finite edge current.
The $p_X$ wave can be realized by applying the uniaxial strain
in the chiral or helical $p$ wave state.

Next, we calculate the edge-induced currents 
due to the edge even-frequency $s$ wave states.
Figures \ref{fig:fig5}(b) and \ref{fig:fig5}(c) are the obtained edge currents 
in the $d+is^\evenf$ wave and $p+s^\evenf$ wave states, respectively. 
The obtained charge current in Fig. \ref{fig:fig5} (b)
is consistent with the Matsumoto-Shiba theory 
\cite{Matsumoto-Shiba-II}.
The parities and edge currents in the 
edge odd- and even-frequency SC states 
are summarized in Table I.

\begin{table}[h]
\caption{\label{tab:table1}%
Parities and edge currents in 
$d+s^\oddf$ $p+is^\oddf$, $d+is^\evenf$, and $p+s^\evenf$ 
wave states for $d=d_{xy}$ and $p=p_x$.
These states satisfy ${\cal M}_x=-1$.
All currents disappear if the phase of the edge gap is shifted by $\pi/2$.
No currents flow for $d=d_{x^2-y^2}$ and $p=p_y$ because ${\cal M}_x=+1$.
}
\begin{ruledtabular}
\begin{tabular}{c c c c c}
SC state & Time-reversal & Spin exchange & $J_x^{\rm C}$ & $J_x^{{\rm S}z}$  
\\ \hline 
$d+s^\oddf$ & $+$ & $-$ & 0 & non zero
\\
$p+is^\oddf$ &$-$ & $+$ & non zero & 0
\\
$d+is^\evenf$ & $-$ & $+$ & non zero & 0
\\
$p+s^\evenf$ & $+$ & $-$ & 0& non zero 
\\
\end{tabular}
\end{ruledtabular}
\end{table}

\section{relationship between $\phi$ and $\phi^+$}
Finally, we discuss a fundamental open problem 
in the relationship between $\phi$ and $\phi^+$ in the odd-frequency gap function.
In this study, we assume the relationship in Eq. (\ref{eqn:Fom}),
which is directly derived from the Lehmann representation.
This relationship gives the para-Meissner effect,
and therefore, it is not as stable as a bulk SC state.
Nonetheless, the odd-frequency gap function is naturally expected 
as the edge-state of bulk superconductivity.
However, a different non-Hermitian relationship,
${\bar \phi}_{y,y'}^+(k_x,i\e_n)= [\phi_{y',y}(k_x,+i\e_n)]^*$,
proposed in Refs. \cite{Kirkpatrick,Solenov,Kusunose,Kusunose2},
which exhibits the usual Meissner effect,
inevitably induces imaginary spin current 
in the $d+s^\oddf$ wave state, as demonstrated in this study.
Therefore, the Hermitian relationship (\ref{eqn:Fom}) 
should be the true equation.

To determine the edge $s^\oddf$ gap function,
it is beneficial to focus on the anomalous proximity 
effect in a diffusive normal metal (DNM),
where the quasiparticle in the DNM exhibits a zero-energy peak of the LDOS
\cite{Tanaka2004}.
In the absence of the edge $s^\oddf$ gap function,
the odd-frequency singlet $p$ wave is solely induced at the interface;
however, it cannot penetrate into the DNM.
Once the $s^\oddf$ triplet SC state is induced,
it can penetrate into the DNM
and generate the zero-energy peak of the LDOS. 

\section{summary}
We have predicted that an odd-frequency spin-triplet $s$ wave 
gap function emerges at the edge of $d$ wave superconductors,
mediated by the zero-energy SABS-induced ferromagnetic fluctuations.
This prediction is obtained from the analysis of the edge SC gap equation
based on the cluster Hubbard model with a bulk $d$ wave gap.
The predicted odd-frequency $s$ wave gap function is expected to 
be robust against randomness.
The obtained SC state with the TRS
accompanies the spontaneous edge spin current.
The predicted edge spin current in the $d+s^\oddf$ wave state is a
useful signal for detecting the hidden odd-frequency SC gap function.
We also provided decisive validation of
the Hermitian relationship [Eq. (\ref{eqn:Fom})]
of the odd-frequency gap function.
An important future issue is to analyze the electronic states below $T_{{\rm c}s}^\oddf$ 
by considering strong coupling effects, like the self-energy and feedback effects.

We have revealed that SABS-driven spin fluctuations at the edge of the bulk superconductor
induce an exotic edge superconductivity.
The SABS-driven spin fluctuations will also induce exotic edge charge density wave (CDW)
due to the paramagnon interference mechanism \cite{Onari-SCVC,Onari-FeSe,Yamakawa-FeSe,Onari-AFN,Onari-B2g}.
The $d$ wave bond order \cite{Kawaguchi-CDW,Tsuchiizu4,BEDT},
$p$ wave charge current order \cite{Tazai-cLC},
and $p$ wave spin current order \cite{Kontani-sLC} are expected to be realized by the 
paramagnon interference mechanism \cite{Tazai-JPSJ}.
The emergence of an edge-induced exotic CDW is an important future issue.

\acknowledgements
We are grateful to S. Onari and Y. Yamakawa for useful discussions.
This work is supported by Grants-in-Aid for Scientific Research (KAKENHI Grants No.
JP19J21693, No. JP19H05825, No. JP18H01175, No.JP18H01176, No.JP20H00131, and No. JP20H01857) from MEXT of Japan,
Japan-RFBR Bilateral Joint Research
Projects Seminars No. 19-52-50026, and the JSPS
Core-to-Core program ``Oxide Superspin'' international network.

\appendix
\section{Linearized gap equation for the edge-induced triplet states}
In this appendix, we derive the linearized triplet gap equation in the presence of the bulk $d$ wave gap
\cite{Matsubara-PRB2}.
First, we assume that $\Delta_{y,y'}^{d}(k_x)$
and the edge triplet gap
$\phi_{y,y'}(k_x,i\epsilon_n)\equiv\phi^{\uparrow\downarrow}_{y,y'}(k_x,i\epsilon_n)$
are both finite.
We ignore the spin orbit interaction,
so we can set the $\bm d$ vector as $\hat{{\bm d}}=(0,0,\hat{\phi})$.
Then, we define the $2N_y \times 2N_y$
Green's functions $\hat{\mathcal{G}}_{\rm Nam}$
in the bulk+edge SC state as follows:
\begin{align}
&\hat{\mathcal{G}}_{\rm Nam}\equiv
\left(
    \begin{array}{cc}
      \hat{\mathcal{G}}^{\uparrow\uparrow}(k_x,i\epsilon_n)
&     \hat{\mathcal{F}}^{\uparrow\downarrow}(k_x,i\epsilon_n) \\
      \hat{\mathcal{F}^+}^{\uparrow\downarrow}(k_x,i\epsilon_n)
&      -{^{t}\hat{\mathcal{G}}^{\downarrow\downarrow}(-k_x,-i\epsilon_n)}\\
    \end{array}
\right)
\nonumber\\
& \ \ \ =
\left(
    \begin{array}{cc}
      i\epsilon_n-\hat{H}^{0}(k_x) 
&     -\hat{\Delta}^d(k_x)-\hat{\phi}(k_x,i\epsilon_n)  \\
      -\hat{\Delta}^d(k_x)-\hat{\phi}^+(k_x,i\epsilon_n) 
&      i\epsilon_n+{^{t}\hat{H}^{0}}(-k_x)\\
    \end{array}
\right)^{-1}.
\label{eq:green_all}
\end{align}
The equation for the triplet gap
$\phi_{y,y'}(k_x,i\epsilon_n)$ is given by
\begin{align}
\phi_{y,y'}(k_x,i\epsilon_n)
=
T
\sum_{k_x',m}
&
V_{y,y'}(k_x-k_x',i\epsilon_n-i\epsilon_m)
\nonumber\\
\times
&\mathcal{F}_{y,y'}^{\rm triplet}(k_x',i\epsilon_m),
\label{eq:gap_eq_derivation_2}
\end{align}
where
$
\hat{\mathcal{F}}^{\rm triplet}(k_x,i\epsilon_n)
\equiv
\{
\hat{\mathcal{F}}^{\uparrow\downarrow}(k_x,i\epsilon_n)
+
\hat{\mathcal{F}}^{\downarrow\uparrow}(k_x,i\epsilon_n)
\}/2
$
is the triplet part of the anomalous Green's function in the coexisting SC state.
In order to linearize \eqref{eq:gap_eq_derivation_2},
we evaluate $\hat{\mathcal{F}}^{\rm triplet}$ by the first-order perturbation of $\hat{\phi}$ and $\hat{\phi}^+$
to the Green's functions \eqref{eq:sc-4}.
Since $\hat{F}_d$ satisfies the relation $\hat{F}_d^{\uparrow\downarrow}=-\hat{F}_d^{\downarrow\uparrow}$,
we obtain
$\hat{\mathcal{F}}^{\rm triplet}=
  -\hat{G}_d
      \hat{\phi}
      \hat{\bar{G}}_d
    +\hat{F}_d
      \hat{\phi}^{+}
      \hat{F}_d
$, 
where $\hat{\bar{G}}_d\equiv{^{t}\hat{G}_d}(-k_x,-i\epsilon_n)$.
By substituting it into Eq. \eqref{eq:gap_eq_derivation_2},
we obtain the analytic expression of the linearized triplet gap equation for $\hat{\phi}$ in Fig. \ref{fig:fig1}(e).
The set of Eqs. (\ref{eq:trip_gap_eq1}) and (\ref{eq:trip_gap_eq2}) 
gives the linearized triplet gap equation
in the presence of the bulk $d$ wave gap.
[In Eqs. (\ref{eq:trip_gap_eq1}) and (\ref{eq:trip_gap_eq2}),
the subscript $d$ of $G$ and $F$ is omitted.]
The edge triplet SC state appears when the eigenvalue 
$\lambda^{\rm edge}$ is around unity.

In the main text, we use the 
Hermitian odd-frequency gap $\phi^+(i\e_n)=-[\phi(i\e_n)]^*$,
and obtain the time-reversal-symmetry $d+s^\oddf$ wave state.
We note that the eigenvalue $\lambda^{\rm edge}$ is unchanged
even if one assumes a non-Hermitian relation $\phi^+(i\e_n)=[\phi(i\e_n)]^*$.

In the present study, the Hermitian odd-frequency gap relation 
gives the finite charge or spin current unless the parity of 
${\cal{M}}_x$ is even.
On the other hand, the non-Hermitian odd-frequency gap relation 
leads to unphysical imaginary currents
in the cases of the $d+is^\oddf$ wave and $p+s^\oddf$ wave states.

\section{$k_x$ dependence of $s^{\oddf}$ gap}
\begin{figure}[htb]
\includegraphics[width=.70\linewidth]{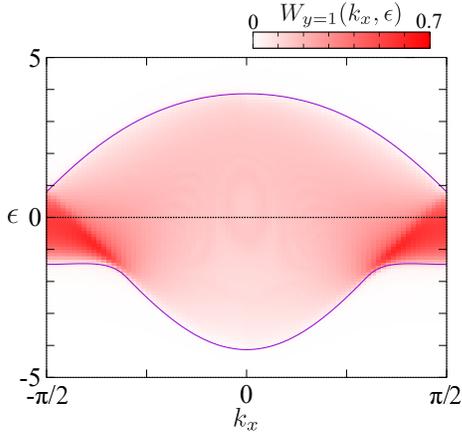}
\caption{
Weight of the edge layer state in the normal state; $W_{y=1}(k_x,\e)$.
}
\label{fig:DOS}
\end{figure}
Figure \ref{fig:DOS} shows the weight of the edge layer state ($y=1$)
in the present cluster tight-binding model without $\Delta^d$.
It is given as
$W_y(k_x,\e)=\sum_{b}\delta(E_{b,k_x}-\e)|U(y,b,k_x)|^2$,
where $E_{b,k_x}$ is the $b$-th band energy at $k_x$ measured from $\mu$ 
and $U(y,b,k_x)$ is the unitary matrix.
Note that the relation $D_y(\e)= \sum_{k_x}W_y(k_x,\e)$ holds.
Since the edge weight is large for $|k_x|\sim \pi/2$,
the magnitude of the $s^\oddf$ wave gap function 
in Fig. 2(a) is large for $|k_x|\sim \pi/2$.

\section{Coherence length of the bulk $d$ wave gap}
\begin{figure}[htb]
\includegraphics[width=.50\linewidth]{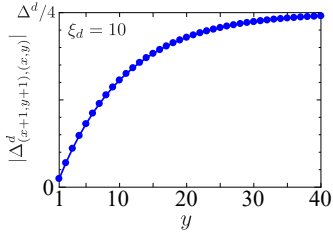}
\caption{
$|\Delta_{i,j}^d|$ for $i=(x,y)$ and $j=(x+1,y+1)$ for $\xi_d=10$.
}
\label{fig:xi}
\end{figure}

The $d$ wave gap function in the Hamiltonian is given as
$\Delta_{i,j}^d=(\Delta^d/4)(\delta_{\bm{r}_i-\bm{r}_j,\pm \hat{\bm{X}}}
-\delta_{\bm{r}_i-\bm{r}_j,\pm \hat{\bm{Y}}})$.
Near the edge layer ($y=1$), 
$\Delta^d_{i,j}$ should be suppressed 
if the $y$-components of the sites $i$ and $j$, $y_i$ and $y_j$,
are smaller than the coherence length $\xi_d=10$.
In order to reproduce this suppression,
we multiply $\Delta_{i,j}^d$ in the Hamiltonian 
by the decay factor $\{1-\exp[(y_i+y_j-2)/2\xi_d]\}$
\cite{Matsubara-PRB2}.
In the main text, we set the coherence length $\xi_d=10$,
and then $|\Delta_{i,j}^d|$ for $i=(x,y)$ and $j=(x+1,y+1)$ 
is given in Fig. \ref{fig:xi}.
From the experimental results \cite{cuprate_coherence_3,cuprate_coherence_4,cuprate_coherence_5,cuprate_lattice_1},
the coherence length in the $a$-$b$ plane of YBCO is 1nm for $T\ll T_{{\rm c}d}$.
Therefore, $\xi_d=10$ is a reasonable value.

\begin{figure}[htb]
\includegraphics[width=.99\linewidth]{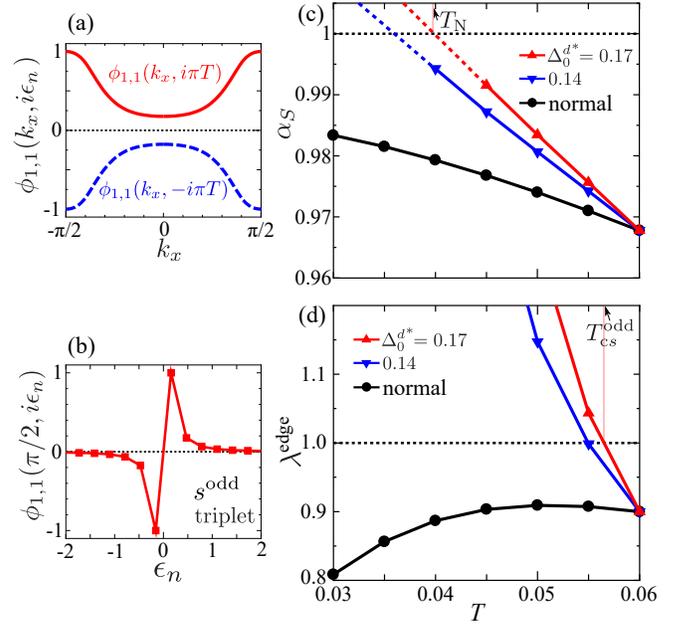}
\caption{
Odd-frequency gap functions obtained by the modified FLEX theory for $U=2.8$.
(a)(b) Obtained $s^\oddf$ wave triplet gap at edge:
(a) $\phi_{1,1}(k_x,\pm i\pi T)$ and (b) $\phi_{1,1}(k_x=\pi/2,i\e_n)$
in case of ${\Delta_0^d}^*=0.17$ at $T=0.05$.
(c)(d) $T$-dependences of (c) the Stoner factor $\a_S$ and
(d) the eigenvalue $\lambda^{\rm edge}$ 
for the $s^\oddf$ wave state.
Here, the bulk $d$ wave SC gap appears at $T_{{\rm c}d}=0.06$.
$2{\Delta_0^d}^*/T_{{\rm c}d}= 4.7$ and 5.6 for ${\Delta_0^d}^*=0.14$ and 0.17,
respectively.
The edge $s^\oddf$ wave gap is obtained for 
$\a_S \gtrsim 0.968$ at $T=T_{{\rm c}d}$.
}
\label{fig:figS1}
\end{figure}

\section{Analysis by modified FLEX approximation}
In the main text, we calculated the 
$y,y'$-dependence of the pairing interaction 
$V_{y,y'}(k_x,i\w_n)$
using the site-dependent RPA theory.
Here, we calculate $V_{y,y'}(k_x,i\w_n)$
using the modified fluctuation-exchange (FLEX) approximation,
in order to study the effect of the self-energy effect,
by following our previous study \cite{Matsubara-PRB1}.
We set $U=2.8$ hereafter.

Figures \ref{fig:figS1}(a) and \ref{fig:figS1}(b) show the 
$q_x$ and $\w_n$ dependences of the odd-frequency $s^\oddf$ wave gap 
at $T=0.05$, respectively.
By setting $\Delta_0^d=0.24$ ($0.20$) in the Hamiltonian,
the normalized $d$ wave gap is obtained as ${\Delta_0^d}^*=0.17$ ($0.14$)
due to the self-energy in the FLEX approximation 
\cite{Matsubara-PRB2}.
The obtained results are similar to those in Figs. 2(a) and 2(b) 
in the main text given by the RPA.

Figures \ref{fig:figS1}(c) and \ref{fig:figS1}(d) exhibit the obtained 
spin Stoner factor $\a_S$ and the eigenvalue $\lambda^{\rm edge}$
as functions of $T$, respectively.
In the normal state (${\Delta_0^d}^*=0$), 
$\a_S$ moderately increases at low temperatures.
In contrast, $\lambda^{\rm edge}$ decreases at low $T$ 
since the pairing interaction for the odd-frequency SC gap
is proportional to $T\chi^s(\q_x,0)$.
In contrast, in the presence of the $d$ wave gap ${\Delta_0^d}^*$,
$\a_S$ rapidly increases due to the 
huge zero-energy surface Andreev bound state (SABS) peak.
Therefore, $\lambda^{\rm edge}$ rapidly increases
owing to the SABS-induced magnetic criticality
\cite{Matsubara-PRB2}.
These results are similar to those in Figs. 2(c) and 2(d) 
in the main text.

Thus, the SABS-driven odd-frequency SC state 
is naturally obtained at the edge of $d$ wave superconductors,
even if the self-energy effect is taken into account 
based on the modified FLEX theory.



\begin{thebibliography}{99}

\bibitem{Buchholtz}
L. J. Buchholtz and G. Zwicknagl. Phys. Rev. B, {\bf 23}, 5788 (1981).

\bibitem{Nagai}
J.Hara and K.Nagai, Prog. Theor. Phys. {\bf 76}, 1237 (1986).


\bibitem{Hu-ZBCP}
C. R. Hu, Phys. Rev. Lett. {\bf 72}, 1526 (1994).

\bibitem{Tanaka-ZBCP}
Y. Tanaka and S. Kashiwaya, Phys. Rev. Lett. {\bf 74}, 3451 (1995).

\bibitem{Kashiwaya-junction}
S. Kashiwaya, Y. Tanaka, M. Koyanagi, K. Kajimura, Phys. Rev. B {\bf 53}, 2667 (1996).

\bibitem{Nagato}
Y. Nagato and K. Nagai, Phys. Rev. B {\bf 51}, 16254 (1995).

\bibitem{Kashiwaya-ZBCP}
S. Kashiwaya and Y. Tanaka, Rep. Prog. Phys. {\bf 63}, 1641 (2000).

\bibitem{Sato_ABS_topo}
M. Sato, Y. Tanaka, K. Yada, and T. Yokoyama, Phys. Rev. B {\bf 83}, 224511 (2011).


\bibitem{Kashiwaya-ZBCP-2}
S. Kashiwaya, Y. Tanaka, M. Koyanagi, H. Takashima, and K. Kajimura, Phys. Rev. B {\bf 51}, 1350 (1995).

\bibitem{Alff}
L. Alff, H. Takashima, S. Kashiwaya, N. Terada, H. Ihara, Y.
Tanaka, M. Koyanagi, and K. Kajimura, 
Phys. Rev. B {\bf 55}, R14757 (1997).

\bibitem{Wei-ZBCP}
J. Y. T. Wei, N. -C. Yeh, D. F. Garrigus, and M. Strasik, Phys. Rev. Lett. {\bf 81}, 2542 (1998).

\bibitem{Geek-ZBCP}
J. Geerk, X. X. Xi, and G. Linker, Z. Phys. B {\bf 73}, 329 (1988).

\bibitem{Matsumoto-Shiba-I}
M. Matsumoto and H. Shiba, J. Phys. Soc. Jpn. {\bf 64}, 3384 (1995).

\bibitem{Matsumoto-Shiba-II}
M. Matsumoto and H. Shiba, J. Phys. Soc. Jpn. {\bf 64}, 4867 (1995).


\bibitem{Kuboki}
K. Kuboki and M. Sigrist, J. Phys. Soc. Jpn. {\bf 67}, 2873 (1998).

\bibitem{Fogelstrom}
M Fogelstr\"{o}m, D. Rainer, and J. A. Sauls, Phys. Rev. Lett. {\bf 79},
281 (1997). 

\bibitem{Potter}
A. C. Potter and P. A. Lee, Phys. Rev. Lett. {\bf 112}, 117002 (2014).

\bibitem{Matsubara-JPSJ}
S. Matsubara, Y. Yamakawa, and H. Kontani, 
J. Phys. Soc. Jpn. {\bf 87}, 073705 (2018).
\bibitem{Matsubara-PRB1}
S. Matsubara and H. Kontani,
Phys. Rev. B {\bf 101}, 075114 (2020).

\bibitem{Matsubara-PRB2}
S. Matsubara and H. Kontani,
Phys. Rev. B {\bf 101}, 235103 (2020).

\bibitem{Berezinskii}
V. L. Berezinskii, JETP Lett. {\bf 20}, 287 (1974).

\bibitem{Kirkpatrick}
T. R. Kirkpatrick and D. Belitz, Phys. Rev. Lett. {\bf 66}, 1533 (1991);
D. Belitz and T. R. Kirkpatrick,
Phys. Rev. B {\bf 60}, 3485 (1999).

\bibitem{Balatsky-Abrahama}
E. Abrahams, A. Balatsky, D. J. Scalapino, and J. R. Schrieffer,
Phys. Rev. B {\bf 52}, 1271 (1995).

\bibitem{Coleman}
P. Coleman, E. Miranda and A. Tsvelik, 
Phys. Rev. Lett. {\bf 70}, 2960  (1993).

\bibitem{Tanaka-odd-frequency}
Y. Tanaka, M. Sato, and N. Nagaosa, J. Phys. Soc. Jpn. {\bf 81},  011013 (2012).

\bibitem{Bala-rev}
J. Linder and A. V. Balatsky, Rev. Mod. Phys. {\bf 91}, 045005 (2019).

\bibitem{Matsumoto2013}
M. Matsumoto, M. Koga, and H. Kusunose,
J. Phys. Soc. Jpn. {\bf 82}, 034708 (2013).

\bibitem{Fominov}
Y. V. Fominov, Y. Tanaka, Y. Asano, and M.Eschrig, Phys. Rev. B {\bf 91}, 144514 (2015).

\bibitem{Yada}
S. Hoshino, K. Yada, and Y. Tanaka, Phys. Rev. B {\bf 93}, 224511 (2016).

\bibitem{Fuseya}
Y. Fuseya, H. Kohno, and K. Miyake,
J. Phys. Soc. Jpn. {\bf 72}, 2914 (2003).

\bibitem{Solenov}
D. Solenov, I. Martin, and D. Mozyrsky, Phys. Rev. B {\bf 79}, 132502 (2009).

\bibitem{Kusunose}
H. Kusunose, Y. Fuseya, and K. Miyake, J. Phys. Soc. Jpn. {\bf 80}, 054702 (2011).

\bibitem{Kusunose2}
H. Kusunose, M. Matsumoto, and M. Koga,
Phys. Rev. B {\bf 85}, 174528 (2012).


\bibitem{Bulut}
N. Bulut, D. J. Scalapino and S. R. White,
Phys. Rev. B {\bf 47} 14599 (1993).

\bibitem{Vojta}
M. Vojta and E. Dagotto, Phys. Rev. B {\bf 59}, R713 (1999).

\bibitem{Hotta}
T. Hotta, J. Phys. Soc. Jpn. {\bf 78}, 123710 (2009).

\bibitem{Shigeta}
K. Shigeta, S. Onari, K. Yada and Y. Tanaka,
Phys. Rev. B {\bf 79}, 174507 (2009).

\bibitem{Yanagi}
Y. Yanagi, Y. Yamashita, and K. Ueda,
J. Phys. Soc. Jpn. {\bf 81}, (2012) 123701.

\bibitem{Heid}
R. Heid, Z. Phys. B {\bf 99}, 15 (1995).

\bibitem{Hoshino}
S. Hoshino, Phys. Rev. B {\bf 90}, 115154 (2014).


\bibitem{Bergeret_2001}
F. S. Bergeret, A. F. Volkov, and K. B. Efetov, Phys. Rev. Lett. {\bf 86}, 4096 (2001); 
Phys. Rev. B {\bf 64}, 134506 (2001).
%
\bibitem{Bergeret_2005}
F. S. Bergeret, A. F. Volkov, and K. B. Efetov, Rev. Mod. Phys. {\bf 77}, 1321 (2005).
%
\bibitem{Golubov_2002}
Ya. V. Fominov, N. M. Chtchelkatchev, and A. A. Golubov
Phys. Rev. B {\bf 66}, 014507 (2002).
%
\bibitem{Yokoyama_2007}
T. Yokoyama, Y. Tanaka, and A. A. Golubov
Phys. Rev. B {\bf 75}, 134510 (2007).
%
\bibitem{Buzdin_2005}
A. I. Buzdin, Rev. Mod. Phys. {\bf 77}, 935 (2005).
%
\bibitem{Buzdin_2011}
A. I. Buzdin, A. S. Mel’nikov, and N. G. Pugach
Phys. Rev. B {\bf 83}, 144515 (2011).
%
\bibitem{Buzdin_2012}
S. Mironov, A. Mel’nikov, and A. Buzdin
Phys. Rev. Lett. {\bf 109}, 237002 (2012).
%
\bibitem{Linder_2009}
Jacob Linder, Takehito Yokoyama, Asle Sudb{\o}, and Matthias Eschrig
Phys. Rev. Lett. {\bf 102}, 107008 (2009).
%
\bibitem{Halterman_2014}
Mohammad Alidoust, Klaus Halterman, and Jacob Linder
Phys. Rev. B {\bf 89}, 054508 (2014).
%
\bibitem{Eschrig_2003}
M. Eschrig, J. Kopu, J. C. Cuevas, and G. Sch¨on, Phys.
Rev. Lett. 90, 137003 (2003).
%
\bibitem{Asano_2007}
Y. Asano, Y. Tanaka, and A. A. Golubov, Phys. Rev. Lett. {\bf 98}, 107002 (2007).
%
\bibitem{Cayao_2020}
J. Cayao, C. Triola, and A. M. Black-Schaffer, Eur. Phys. J. Special Topics {\bf 229}, 545 (2020).


\bibitem{Tanaka-odd-frequency-2}
Y. Tanaka and A. A. Golubov, Phys. Rev. Lett. {\bf 98}, 037003 (2007).

\bibitem{Tamura}
S. Tamura, S. Hoshino, and Y. Tanaka,
Phys. Rev. B {\bf 99}, 184512 (2019).

\bibitem{Tanaka_2007_1}
Y. Tanaka, A. A. Golubov, S. Kashiwaya, and M. Ueda. Phys. Rev. Lett. {\bf 99}, 037005 (2007).

\bibitem{Tanaka_2007_2}
Y. Tanaka, Y. Tanuma, and A. A. Golubov. Phys. Rev. B {\bf 76}, 054522 (2007).

\bibitem{Tanaka2004}
Y. Tanaka and S. Kashiwaya: Phys. Rev. B {\bf 70}  012507 (2004).

\bibitem{Higashitani2009}
S. Higashitani, Y. Nagato, and K. Nagai, Journal of Low
Temperature Physics 155, 83 (2009).

\bibitem{Higashitani}
 S. Higashitani: J. Phys. Soc. Jpn. {\bf 66}  2556 (1997). 

\bibitem{Tanaka2005R}
 Y. Tanaka, Y. Asano, A. Golubov, and S. Kashiwaya: Phys. Rev. B
{\bf 72} 140503(R) (2005).

\bibitem{Suzuki-Asano1}
S. Suzuki and Y. Asano, Phys. Rev. B {\bf 89}, 184508 (2014). 

\bibitem{Suzuki-Asano2}
S. Suzuki and Y. Asano, Phys. Rev. B {\bf 91}, 214510 (2015). 

\bibitem{Bernardo}
A. Di Bernardo, Z. Salman, X. L. Wang, M. Amado,
M. Egilmez, M. G. Flokstra, A. Suter, S. L. Lee, J. H.
Zhao, T. Prokscha, E. Morenzoni, M. G. Blamire, J. Linder,
and J. W. A. Robinson, Phys. Rev. X 5, 041021
(2015).

\bibitem{Krieger}
J. A. Krieger, A. Pertsova, S. R. Giblin, M. D\"{o}beli,
T. Prokscha, C. W. Schneider, A. Suter, T. Hesjedal,
A. V. Balatsky, and Z. Salman, Phys. Rev. Lett. 125,
026802 (2020).

\bibitem{Moriya}
 T. Moriya and K. Ueda: Adv. Phys. {\bf 49} 555 (2000).

\bibitem{Scalapino}
D. J. Scalapino,
Rev. Mod. Phys. {\bf 84}, 1383 (2012).

\bibitem{Tremblay}
Y. M. Vilk, A. -M. S. Tremblay,
J. Phys. I France {\bf 7}, 1309 (1997).

\bibitem{Chubukov}
A. V. Chubukov, D. Pines, and J. Schmalian,
{\it Superconductivity: Conventional and Unconventional Superconductors},
edited by K.-H. Bennemann and J. B. Ketterson
(Springer, Berlin, 2003).

\bibitem{Kontani-rev}
H. Kontani, Rep. Prog. Phys. {\bf 71} (2008) 026501.


\bibitem{Chen}
Y. Chen and C. S. Ting,	Phys. Rev. Lett. {\bf 92}, 077203 (2004).

\bibitem{Kontani-imp}
H. Kontani and M. Ohno,
Phys. Rev. B {\bf 74}, 014406 (2006);
J. Magn. Magn. Mater. {\bf 310}, 483 (2007).

\bibitem{Alloul99-2}
P. Mendels, J. Bobroff, G. Collin, H. Alloul, M. Gabay, 
J.F. Marucco, N. Blanchard and B. Grenier,
Europhys. Lett. {\bf 46} 678 (1999).

\bibitem{Ishida96}
K. Ishida, Y. Kitaoka, K. Yamazoe, K. Asayama, and Y. Yamada,
Phys. Rev. Lett. {\bf 76} 531 (1996).


\bibitem{Yamakawa-CDW}
Y. Yamakawa and H. Kontani, Phys. Rev. Lett. {\bf 114}, 257001 (2015).

\bibitem{Kawaguchi-CDW}
K. Kawaguchi, Y. Yamakawa, M. Tsuchiizu, and H. Kontani,
J. Phys. Soc. Jpn. {\bf 86}, 063707 (2017).

\bibitem{Tsuchiizu4}
M. Tsuchiizu, K. Kawaguchi, Y. Yamakawa, and H. Kontani, 
Phys. Rev. B {\bf 97}, 165131 (2018).

\bibitem{cuprate_coherence_1}
D. S. Inosov, J. T. Park, A. Charnukha, Yuan Li, A. V. Boris, B. Keimer, and V. Hinkov
Phys. Rev. B {\bf 83}, 214520 (2011).

\bibitem{cuprate_coherence_2}
{\O}ystein Fischer, Martin Kugler, Ivan Maggio-Aprile, Christophe Berthod, and Christoph Renner
Rev. Mod. Phys. {\bf 79}, 353 (2007).

\bibitem{H.Kashiwaya}
H. Kashiwaya, S. Kashiwaya, B. Prijamboedi, A. Sawa, I. Kurosawa, Y. Tanaka, and I. Iguchi
Phys. Rev. B 70, 094501 (2004).

\bibitem{Bouscher}
S. Bouscher, Z. Kang, K. Balasubramanian, D. Panna,
P. Yu, X. Chen, and A. Hayat, 32, 475502 (2020).

\bibitem{PALee}
A. C. Durst and P. A. Lee, Phys. Rev. B {\bf 62} 1270 (2000).

\bibitem{Saadaoui}
H. Saadaoui, Z. Salman, T. Prokscha, A. Suter, H. Huhti-
nen, P. Paturi, and E. Morenzoni, Phys. Rev. B {\bf 88}, 180501
(2013).

\bibitem{Onari-SCVC}
S. Onari and H. Kontani, Phys. Rev. Lett. {\bf 109}, 137001 (2012).
\bibitem{Onari-FeSe}
S. Onari, Y. Yamakawa, and H. Kontani, Phys. Rev. Lett. {\bf 116}, 227001 (2016).
\bibitem{Yamakawa-FeSe}
Y. Yamakawa, S. Onari, and H. Kontani, Phys. Rev. X {\bf 6}, 021032 (2016).
\bibitem{Onari-AFN}
S. Onari and H. Kontani, Phys. Rev. Res {\bf 2}, 042005(R) (2020).
\bibitem{Onari-B2g}
S. Onari and H. Kontani, Phys. Rev. B {\bf 100}, 020507(R) (2019).
\bibitem{BEDT}
R. Tazai, Y. Yamakawa, M. Tsuchiizu, and H. Kontani, Phys. Rev. Res. {\bf 3}, L022014 (2021).
\bibitem{Tazai-cLC}
R. Tazai, Y. Yamakawa, and H. Kontani, Phys. Rev. B {\bf 103}, L161112 (2021).
\bibitem{Kontani-sLC}
H. Kontani, Y. Yamakawa, R. Tazai, and S. Onari, Phys. Rev. Res. {\bf 3}, 013127 (2021).
\bibitem{Tazai-JPSJ}
R. Tazai, Y. Yamakawa, M. Tsuchiizu, and H. Kontani, arXiv:2105.01872.

\bibitem{cuprate_coherence_3}
Y. Matsuda, T. Hirai, S. Komiyama, T. Terashima, Y. Bando, K. Iijima, K. Yamamoto, and K. Hirata
Phys. Rev. B {\bf 40}, 5176 (1989).
%
\bibitem{cuprate_coherence_4}
K. Semba, A. Matsuda, and T. Ishii
Phys. Rev. B {\bf 49}, 10043 (1994).
%
\bibitem{cuprate_coherence_5}
K. Tomimoto, I. Terasaki, A. I. Rykov, T. Mimura, and S. Tajima
Phys. Rev. B {\bf 60}, 114 (1999).
%
%
\bibitem{cuprate_lattice_1}
F. Izumi, H. Asano, T. Ishigaki, A. Ono, and F. P. Okamura
Jpn. J. Appl. Phys. {\bf 26}, L611 (1987).

\end{thebibliography}
\end{document}